\documentclass{iau}
\usepackage{graphicx}
\usepackage{amsmath}
\usepackage{ifthen,hyperref}

\title[Bayesian CMB foreground separation] %% give here short title %%
{Bayesian CMB foreground separation with a correlated log-normal model}

\author[Niels~Oppermann \& Torsten~A.~En{\ss}lin]   %% give here short author list %%
{Niels~Oppermann$^1$
 \and Torsten~A.~En{\ss}lin$^2$}

\affiliation{$^1$Canadian Instittute for Theoretical Astrophysics, University of Toronto, \\ 60 St. George Street, Toronto, ON, M5S 3H8, Canada \\ email: {\tt niels@cita.utoronto.ca} \\[\affilskip]
$^2$Max Planck Institute for Astrophysics, \\ Karl-Schwarzschild-Stra{\ss}e 1, 85748 Garching, Germany \\ email: {\tt ensslin@mpa-garching.mpg.de}}

\pubyear{2014}
\volume{306}  %% insert here IAU Symposium No.
\pagerange{???--???}
\jname{Statistical Challenges in 21st Century Cosmology}
\editors{A.~Heavens, J.-L.~Starck \& A.~Krone-Martins, eds.}

\begin{document}

\maketitle

\begin{abstract}
The extraction of foreground and CMB maps from multi-frequency observations relies mostly on the different frequency behavior of the different components. Existing Bayesian methods additionally make use of a Gaussian prior for the CMB whose correlation structure is described by an unknown angular power spectrum. We argue for the natural extension of this by using non-trivial priors also for the foreground components. Focusing on diffuse Galactic foregrounds, we propose a log-normal model including unknown spatial correlations within each component and cross-correlations between the different foreground components. We present case studies at low resolution that demonstrate the superior performance of this model when compared to an analysis with flat priors for all components.
\end{abstract}

\firstsection % if your document starts with a section,
\section{Introduction}

The separation of the CMB and the various diffuse foreground components relies mostly on their different frequency dependence. Bayesian methods are especially useful for this, since all their assumptions are made explicitly and the remaining uncertainties are easily quantifiable (see, e.g., \cite{planck-2013XII}). Established Bayesian methods infer a set of parameters explicitly describing the frequency spectra of a set of physical foreground components under the assumption of non-trivial priors for these parameters (\cite{eriksen-2006}). On the other hand, the priors for the parameters describing the spatial dependence of the emission components, i.e., the pixel values, are usually assumed to be flat (\cite{planck-2013XII}), with the exception of the CMB component, for which an isotropic Gaussian prior is sometimes assumed (\cite{eriksen-2008}). Here, we propose a natural extension, namely the inclusion of non-trivial spatial priors for the foreground components. This will enable us to allow for spatial correlations both within each component and across components. Specifically, we suggest the use of isotropic log-normal priors for the diffuse Galactic components. For simplicity, we will assume in the following that the frequency dependence of every physical component is known.

\section{Data and signal model}

We can relate the observed frequency maps $d$ to maps of the emission components $\phi$ via the linear model
\begin{equation}
	\label{eq:datamodel}
	d = R\,\phi + n.
\end{equation}
Here, the linear operator $R$ mixes the components into several frequency maps according to their frequency spectra. It can in principle also mix the information from different points in the component maps into different pixels of the observed maps, e.g., via a beam convolution. The last term in Eq.~\eqref{eq:datamodel} is an additive noise term, for which we assume Gaussian statistics with zero mean and a known covariance matrix. Thus, the likelihood $\mathcal{P}{(d|\phi)}$ is fully determined.

To find the posterior, $\mathcal{P}{(\phi|d)}$, which is the probability distribution that we are interested in, we need to augment the likelihood with a suitable prior distribution $\mathcal{P}{(\phi)}$. For this we suggest to model the CMB as an isotropic Gaussian random field and the foreground components as isotropic log-normal fields, i.e., we define $s^{(0)}_{\hat{n}} = \phi^{(0)}_{\hat{n}}/\bar{\phi}^{(0)}$ for the CMB component and $s^{(\alpha)}_{\hat{n}} =\log{\left(\phi^{(\alpha)}_{\hat{n}}/\bar{\phi}^{(\alpha)}\right)}$ for each of the foreground components, where the barred quantities are suitable dimensional normalization constants. We then use an isotropic Gaussian prior for the transformed components, $s$, that is described by a zero mean and an isotropic covariance. The latter can be written in the spherical harmonic basis as
\begin{equation}
	S^{(\alpha,\alpha')}_{(\ell,m),(\ell',m')} = \left< s^{(\alpha)}_{(\ell,m)} \, s^{(\alpha')*}_{(\ell',m')} \right>_{(s)} = \delta_{\ell \ell'} \, \delta_{m m'} \, C^{(\alpha,\alpha')}_\ell,
\end{equation}
where the asterisk denotes complex conjugation. Here, the auto-spectra, $C^{(\alpha,\alpha)}_\ell$, describe the angular correlation structure within the components and the cross-spectra, $C^{(\alpha,\alpha')}_\ell$ for $\alpha \neq \alpha'$, describe the correlations between different components. Both types of quantities are clearly non-zero in nature and accounting for them will allow us to transfer some of the information from one pixel into the reconstruction of other pixels and from one component into the reconstruction of other components.

To complete the model, we need to specify a prior for the unknown auto- and cross-spectra. For this we use an inverse-Wishart distribution as the natural generalization of the inverse-Gamma prior commonly used for unknown angular power spectra (\cite{o'hagan-1994}). With this, the angular spectra can be marginalized and the posterior $\mathcal{P}{(s|d)}$ calculated.

A simpler way to account for correlations would be to model all emission components as Gaussian random fields. However, the log-normal model has three advantages that make it more suited to describe the Galactic foregrounds: The foreground intensities will automatically be positive, their fluctuations will easily be able to vary over orders of magnitude, such as observed between the Galactic plane and the Galactic halo, and the logarithmic model will suppress the anisotropies that are definitely present in the sky to some degree.

\section{Reconstruction algorithm and test cases}

The posterior probability distribution $\mathcal{P}{(s|d)}$ can be calculated analytically. However, it is non-Gaussian. One way to explore this probability distribution would be to draw samples from it using a Monte Carlo algorithm. To minimize computational costs, we pursue a different route here.

Following \cite{oppermann-2013}, we instead apply an iterative scheme that finds the Gaussian distribution that best approximates (in a minimum Kullback-Leibler distance sense) the true posterior, described by a mean and a covariance. The details of the resulting algorithm are beyond the scope of these proceedings and will be described in a forthcoming publication.

\begin{figure}
%	\vspace*{-0.3 cm}
	\begin{center}
		\includegraphics[width=0.19\textwidth]{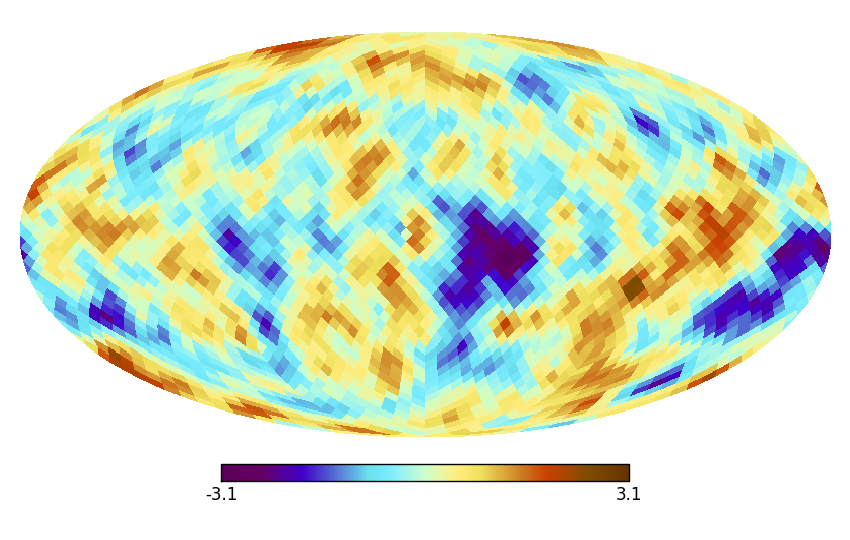}
		\includegraphics[width=0.19\textwidth]{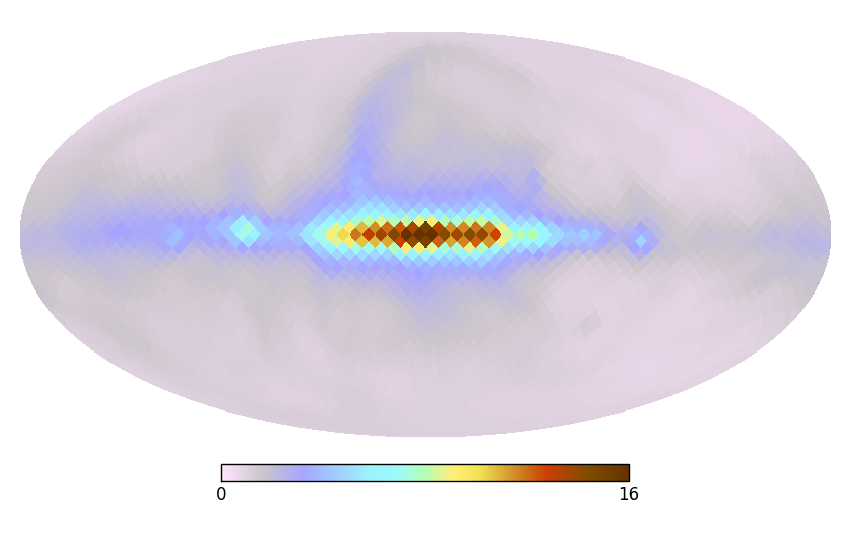}
		\includegraphics[width=0.19\textwidth]{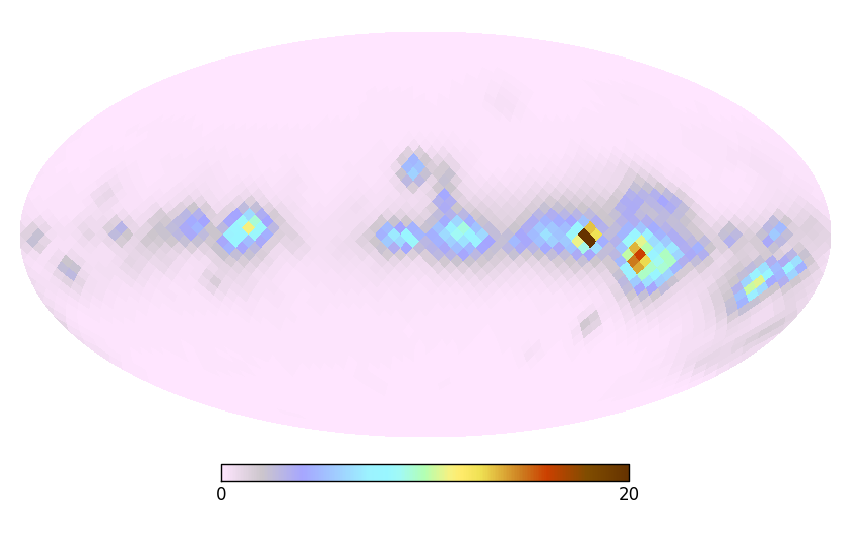}
		\includegraphics[width=0.19\textwidth]{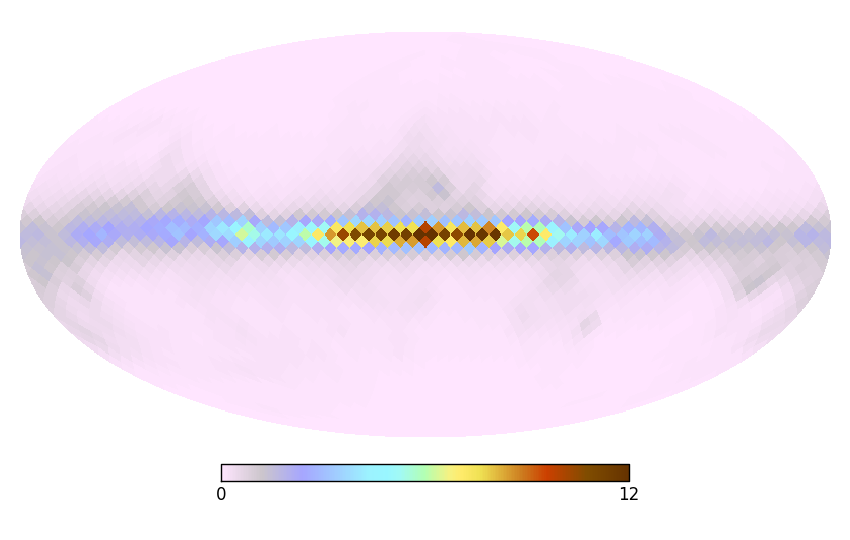} \\
		\includegraphics[width=0.19\textwidth]{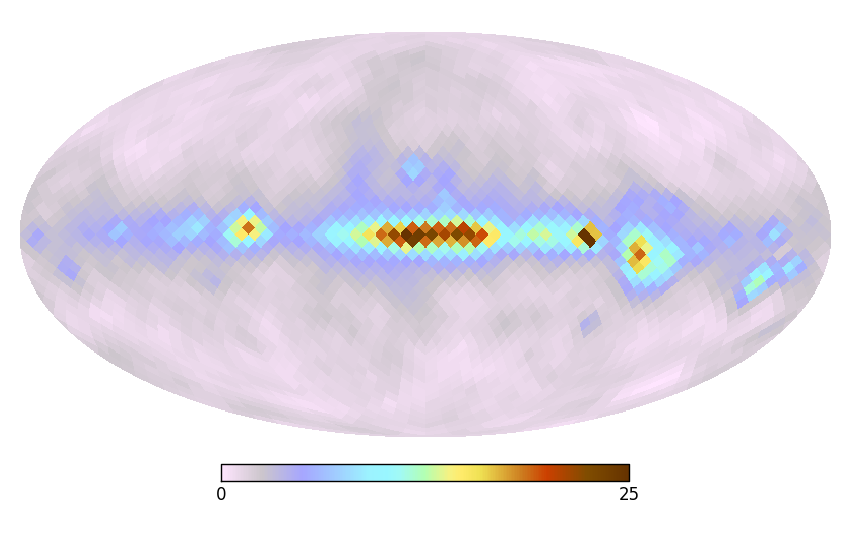}
		\includegraphics[width=0.19\textwidth]{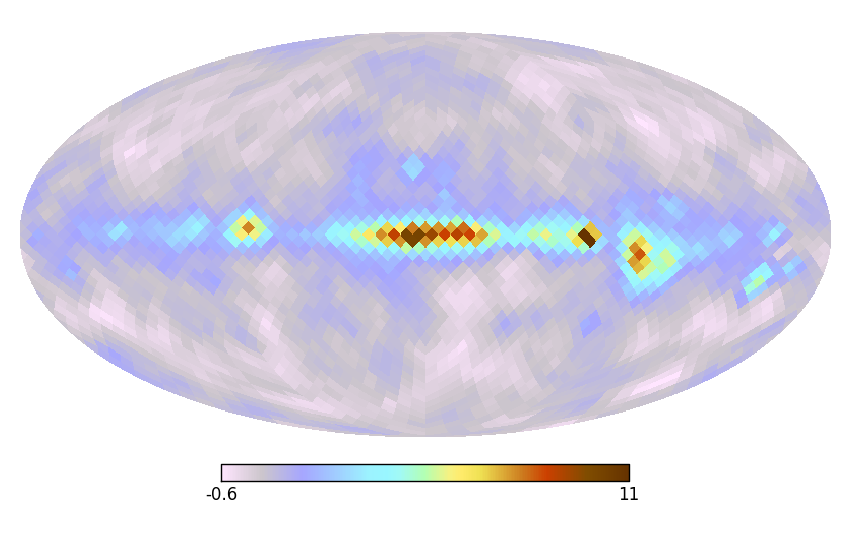}
		\includegraphics[width=0.19\textwidth]{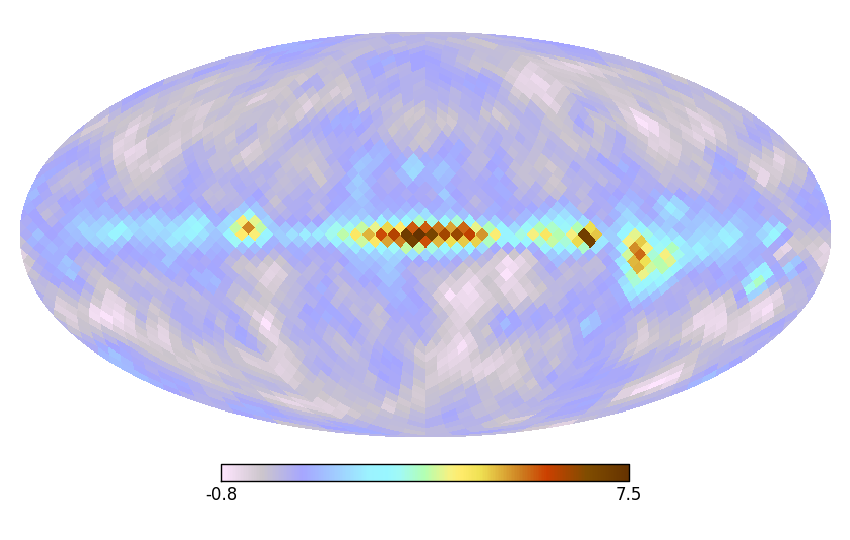}
		\includegraphics[width=0.19\textwidth]{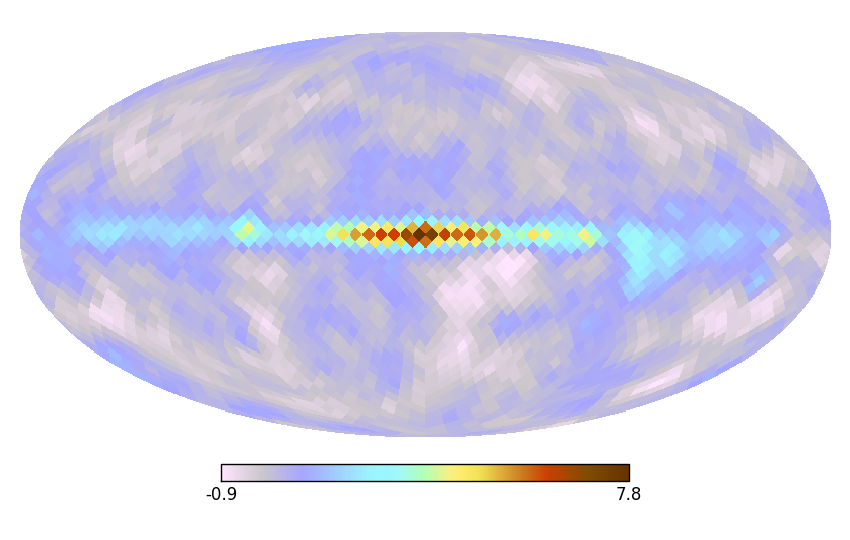}
		\includegraphics[width=0.19\textwidth]{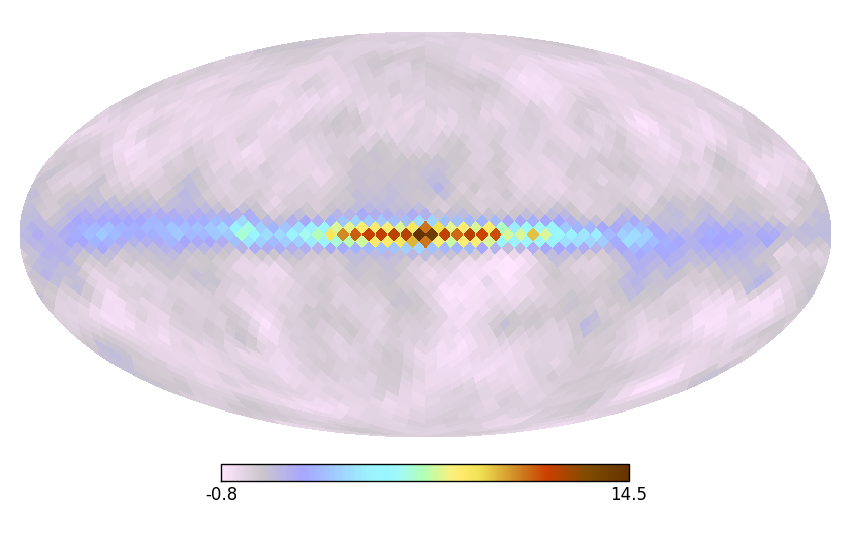} \\
		\includegraphics[width=0.19\textwidth]{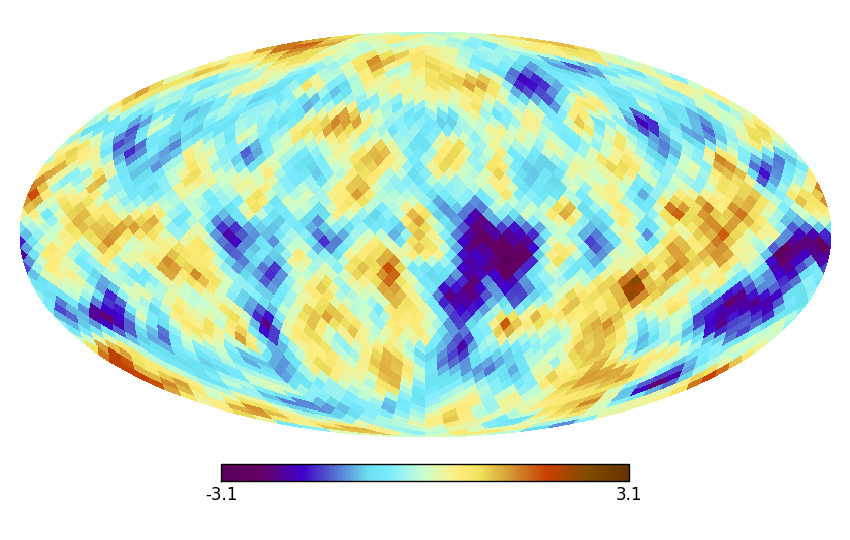}
		\includegraphics[width=0.19\textwidth]{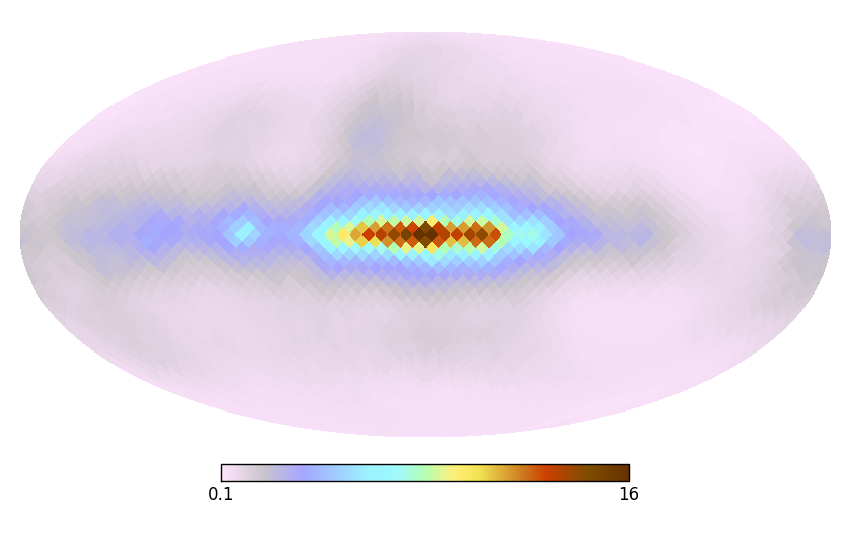}
		\includegraphics[width=0.19\textwidth]{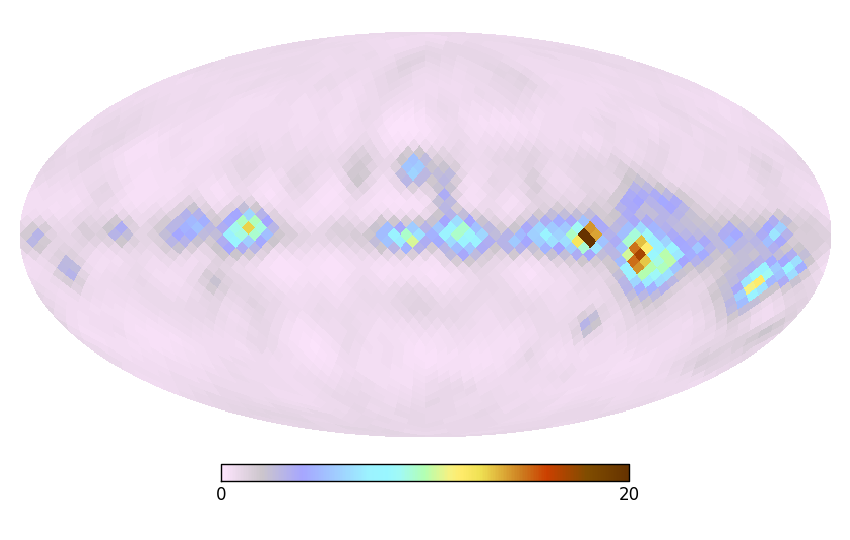}
		\includegraphics[width=0.19\textwidth]{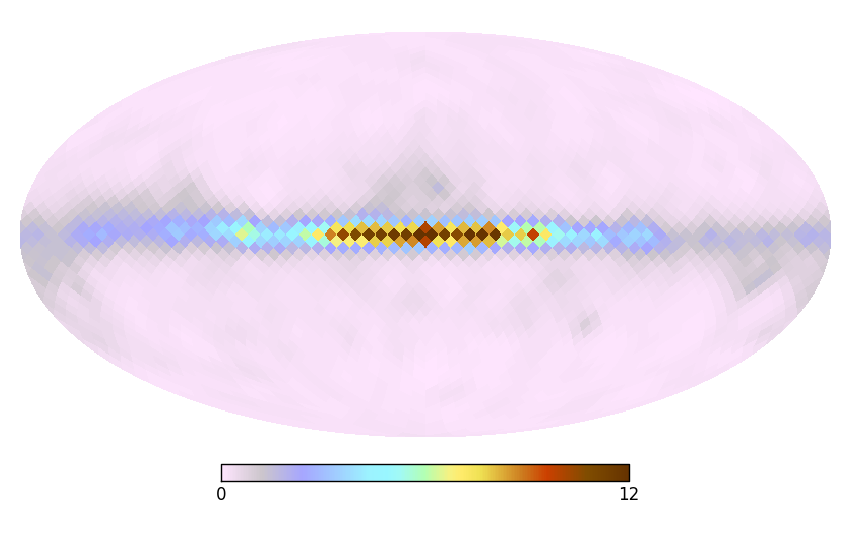} \\
		\includegraphics[width=0.19\textwidth]{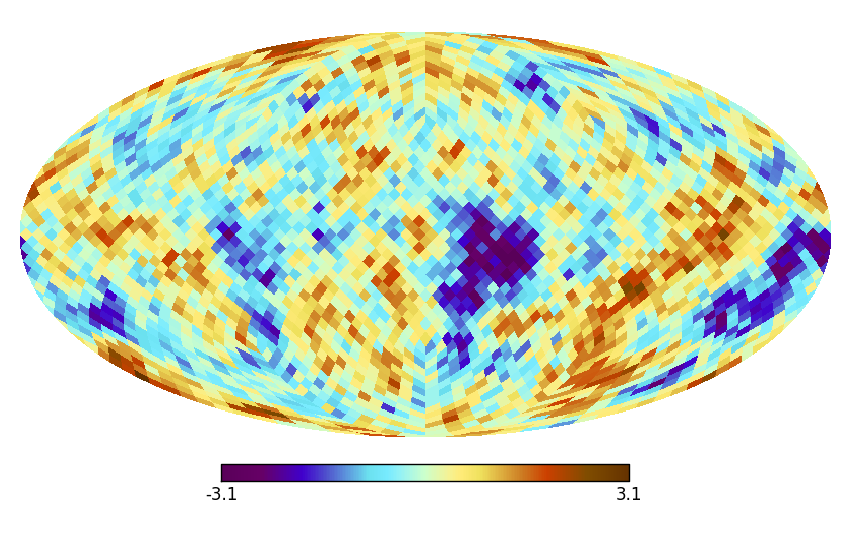}
		\includegraphics[width=0.19\textwidth]{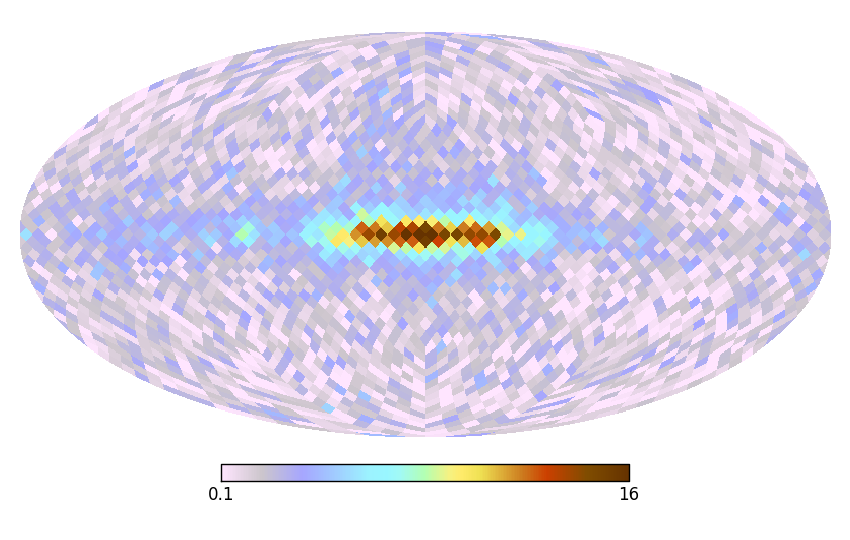}
		\includegraphics[width=0.19\textwidth]{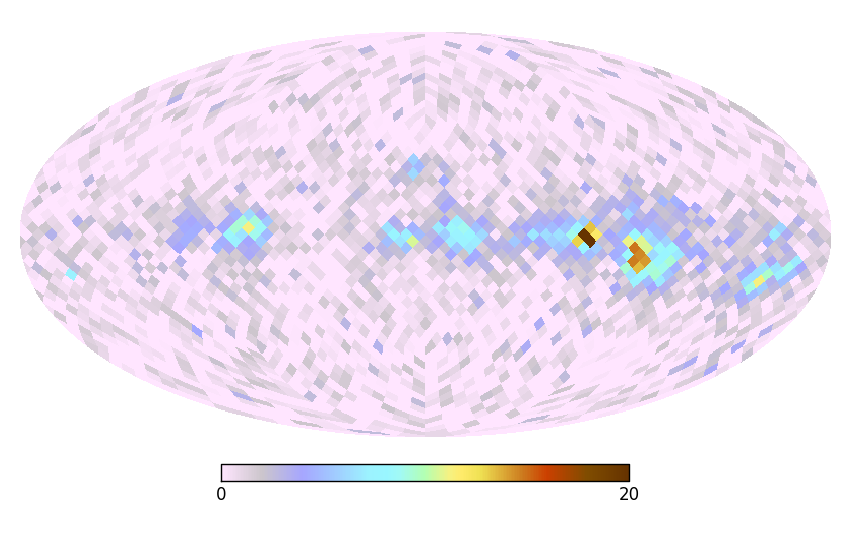}
		\includegraphics[width=0.19\textwidth]{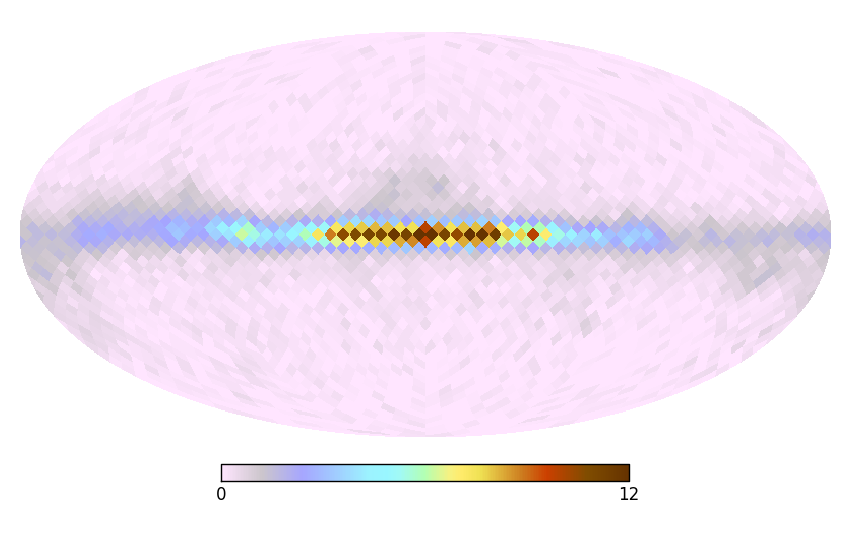}
		\vspace*{-0.3 cm}
		\caption{Simulated sky and its reconstructions. The top row shows, from left to right, a CMB realization (based on the \texttt{Commander-Ruler} samples of \cite{planck-2013XII}), a synchrotron map (based on \cite{haslam-1982}), a free-free map (based on \cite{bennett-2013}), and a dust map (taken from the \texttt{Commander-Ruler} results of \cite{planck-2013XII}). The second row shows simulated frequency maps, containing an isotropic noise contribution. Typical frequency spectra for the different components have been assumed. The third and fourth rows show the reconstructions of the four simulated components using the algorithm presented here and a maximum likelihood calculation, respectively.}
		\label{fig:testreal}
	\end{center}
\end{figure}

In Fig.~\ref{fig:testreal}, we show a simplistic low-resolution simulation of a CMB sky and three foregrounds meant to mimic synchrotron radiation, free-free radiation, and thermal dust radiation, along with a simulated data set generated from these maps. The third and fourth rows of the figure show the posterior mean reconstruction using our approximative algorithm and a maximum likelihood reconstruction, which corresponds to the posterior mean under the assumption of flat priors for the fields $\phi$, respectively. This implementation has made use of the \texttt{NIFTy} package (\cite{selig-2013}) and the \texttt{HEALPix} package (\cite{gorski-2005}).

\section{Conclusion}

Our test case demonstrates that the correlated log-normal prior model is not only well motivated as a good approximation of reality, but also produces superior results in practice, compared to an analysis using flat priors. This is true even if reality, like the test case presented here, is not precisely described by isotropic log-normal distributions.

\end{document}